\documentclass{article}

\usepackage{microtype}
\usepackage{graphicx}
\usepackage{subcaption}
\usepackage{booktabs}
\usepackage{hyperref}

\usepackage[accepted]{icml2026}

\usepackage{xspace}
\usepackage{amsmath}
\usepackage{amssymb}
\usepackage{mathtools}
\usepackage{amsthm}

\newcommand{\parabf}[1]{\noindent\textbf{#1}}

\newcommand{\paraf}[1]{\noindent\textbf{#1}}
\newcommand{\cut}[1]{}

\newcommand{\sysname}{Valve\xspace}

\icmltitlerunning{Valve: Production Online–Offline Inference Colocation with Jointly-Bounded Preemption Latency and Rate}

\begin{document}
\sloppy

\twocolumn[
  \icmltitle{\sysname: Production Online–Offline Inference Colocation with Jointly-Bounded Preemption Latency and Rate}

  \begin{icmlauthorlist}
    \icmlauthor{Fangyue Liu\textsuperscript{*}}{pku,tencent}
    \icmlauthor{Hua Liu\textsuperscript{*}}{tencent}
    \icmlauthor{Xinyuan Lyu\textsuperscript{*}}{tencent}
    \icmlauthor{Shuo Ai}{tencent}
    \icmlauthor{Hao Liang}{tencent}
    \icmlauthor{Lingpeng Chen}{tencent}
    \icmlauthor{Ziqian Hu}{tencent}
    \icmlauthor{Chong Zha}{tencent}
    \icmlauthor{Xin Jin}{pku}
    \icmlauthor{Hanmei Luo}{tencent}
    \icmlauthor{Peng Chen}{tencent}
  \end{icmlauthorlist}

  \icmlaffiliation{pku}{Peking University}
  \icmlaffiliation{tencent}{Tencent}

  \icmlkeywords{GPU sharing, Inference, Scheduling}

  \vskip 0.3in
]

% This command creates the footnote in the first column listing the
% affiliations and the copyright notice. Keep empty braces if no special notice.
\printAffiliationsAndNotice{\textsuperscript{*}Equal contribution.}

\begin{abstract}

LLM inference powers latency-critical production services nowadays.
The bursty nature of inference traffic results in over-provisioning,
which in turn leads to resource underutilization.
While online-offline colocation promises to utilize idle capacity,
broad production deployment must overcome two major challenges:
(i) large online interference due to slow or frequent preemptions,
and (ii) extensive frameworks and drivers modifications, to colocate different models and support preemptions.
We present \sysname, a production-friendly colocation system that jointly bounds
\emph{preemption latency} and \emph{preemption rate}.
Specifically, \sysname enables sub-millisecond compute preemption at most once per online request,
and rate-limited sub-layer memory reclamation.
These guaranties are provided by a GPU runtime that combines channel-controlled
compute isolation, page-fault-free memory reclamation, and dynamic memory reservation.
Critically, \sysname is practical to deploy, requiring one line of driver modification
and 20 lines of framework patch.
% \sysname further employs a burst- and multi-GPU-aware scheduler to utilize
% the harvested headroom efficiently.
Deployed on 8,054 GPUs in production, \sysname improves cluster utilization by 34.6\%,
which translates to a 2,170 GPU save.
This efficiency gains is achieved with minimal online interference,
incurring $<5\%$ TTFT increase and $<2\%$ TPOT increase across workloads.

\end{abstract}
\section{Introduction}
\label{sec:introduction}

Large language model (LLM) inference powers a growing set of workloads.
These include latency-critical production services, such as conversational
assistants, code generation, multimodal tasks~\cite{chatgpt,claude, liu2024ii, jimenez2023swe, jain2024livecodebench}.
LLM also serves batch processing workloads, such as document processing, data analysis~\cite{lai2023ds}.
Beyond user-facing applications, inference has also become a building block for
training workflows, including data curation, post-training actor rollouts, and critic scoring~\cite{guo2025deepseek, lee2023rlaif}.

Despite their importance, LLM inference clusters in production still suffer from
low utilization. This is primarily because operators must provision for bursty demand under strict latency SLAs, resulting in significant idle capacity off-peak.
In practice, the burstiness comes from two main sources.
First, online services may experience unpredictable traffic spikes~\cite{xiang2025aegaeon},
which are further amplified by customized or fine-tuned models.
Second, inference in training workflows typically arrives in periodic large batches,
causing volatile GPU memory usage~\cite{zhong2025optimizing, wu2025hybridflow}.
Harvesting idle capacity to improve utilization remains one of
the central challenges in LLM inference deployments.

\begin{figure}[t]
    \centering
    \includegraphics[width=1.0\linewidth]{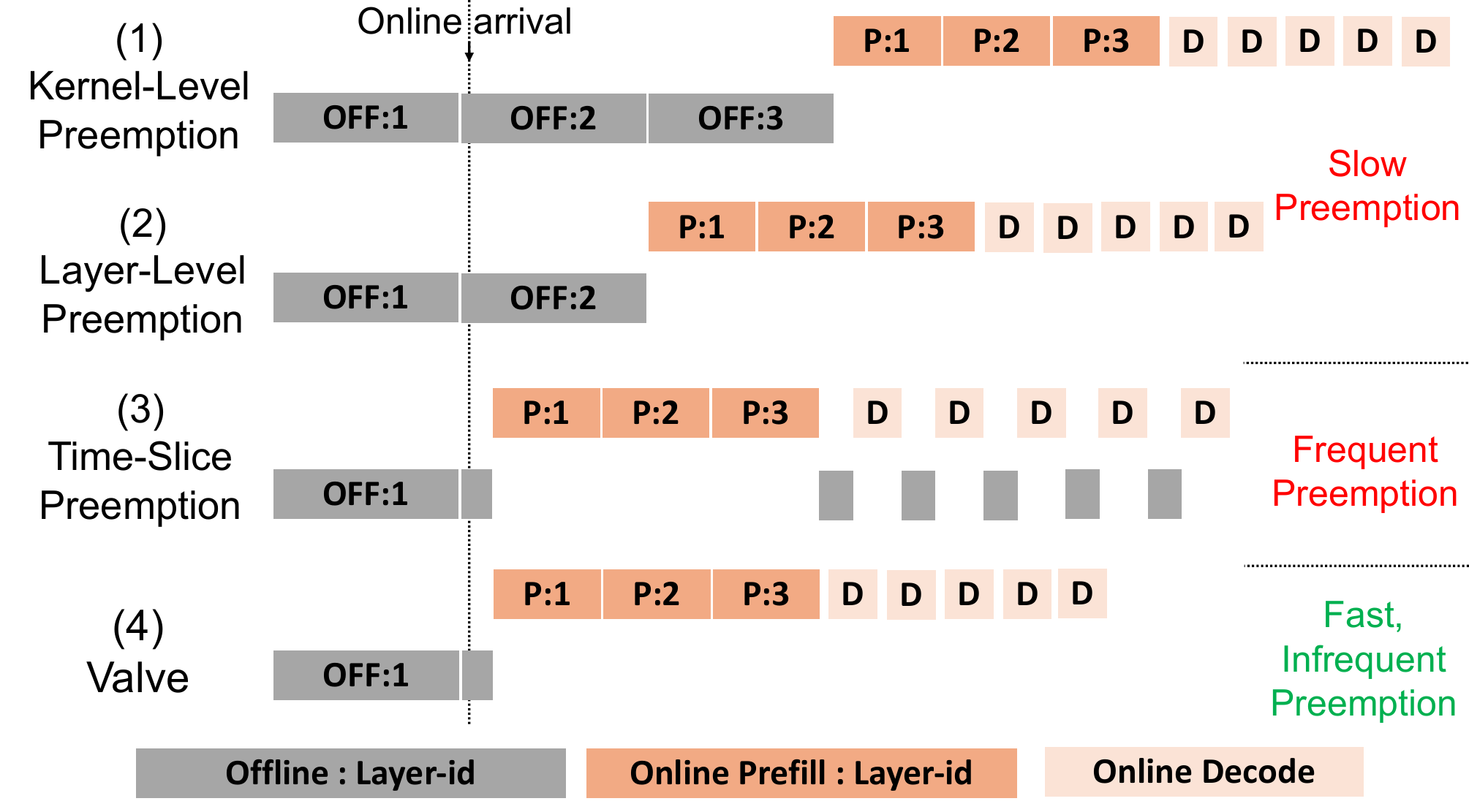}
    \caption{Comparison of online-offline colocation approaches.}
    \label{fig:intro-comparison}
\end{figure}

A promising direction is to colocate latency-critical online serving
with offline inference on the same GPU, which utilizes the idle capacity
to run preemptible offline workloads.
In practice, however, we find two key obstacles that limit broad deployments
of existing approaches~\cite{wu2023transparent, qiao2024conserve, fan2025gpreempt}:
$(i)$ \textit{interference with online workloads} caused by high preemption
latency or frequency, and $(ii)$ \textit{extensive modifications} to GPU drivers or inference frameworks.

Coarse-grained preemption—e.g., kernel-level~\cite{wu2023transparent} (which degrades to iteration-level when CUDA graphs are enabled) and
transformer-layer-level~\cite{qiao2024conserve}—incurs preemption latencies of up to tens of milliseconds.
Immediate preemption~\cite{fan2025gpreempt, cuda_unified_memory} can react quickly, but may trigger preemptions frequently.
Moreover, many existing approaches require extensive framework or driver modifications~\cite{ruan2023nu, qiao2024conserve},
which hinders deployment in production environments.

To address these challenges, we present \sysname, a production-friendly online-offline
colocation system that jointly bounds \emph{preemption latency} and \emph{preemption rate},
while imposing negligible interference with online services. \sysname builds on three key ideas:

(1) \textbf{Channel-controlled compute isolation.} \sysname uses GPU channel control to
\emph{preempt and recover} offline execution within sub-millisecond latency. Combined with \emph{online request lifecycle awareness}, \sysname gates offline kernels outside the lifetime of an online request, ensuring each online request is preempted at most once.

(2) \textbf{Sub-layer memory reclamation with dynamic reservation.}
\sysname reclaims KV cache promptly by coordinating memory reclamation with compute preemption. After preempting compute, \sysname remaps reclaimed pages to a quarantine page and exposes invalidated page IDs to the framework for recomputation, preventing unrecoverable page faults. \sysname further regulates reclamation rate by dynamically reserving memory for online workloads.
% and optionally achieves higher offline throughput via selective reclamation for fewer preempted offline requests.

(3) \textbf{Throughput-aware scheduling.} A burst and multi-GPU aware
scheduler models offline throughput on harvested GPUs
and assigns offline workloads, meeting their SLAs.

We build \sysname as a production-friendly system consisting of a node-level runtime
and a cluster-level scheduler. \sysname requires only a one-line driver
modification and 20-line framework patching, making it easy to deploy in production.
The main contributions of this paper are summarized as follows:
\begin{itemize}
    \item We design a production-friendly runtime that enables
    sub-millisecond compute preemption at most once per online request,
    and sub-layer memory reclamation with a rate-bounded reclamation frequency.
    \item We develop a burst- and multi-GPU-aware scheduler that places offline
    jobs smartly on harvested GPUs to meet throughput SLAs while improving utilization. 
    \item We deploy \sysname in a production cluster with 8,054 GPUs, improving
    average utilization by 34.6\%, which translates to saving 2,170 GPUs. Across workloads, \sysname incurs $<5\%$ TTFT increase and $<2\%$ TPOT increase.
\end{itemize}
\section{Background}
\label{sec:background}

% In this section, we first analyze the low GPU utilization by
% examining the burstiness and SLAs of production workloads.
% Second, we derive key requirements for production-friendly online-offline inference colocation.
% Third, we review existing online-offline colocation systems, discussing their limitations in production.
% Finally, we discuss the main challenges in designing a production-friendly inference colocation
% meeting the key requirements.

We first analyze why GPU utilization is low in production by characterizing workload burstiness
and SLA requirements. We then derive key requirements for production-friendly online--offline
inference colocation, review existing systems and their limitations, and outline the main challenges.

\subsection{Low GPU Utilization of Production Workloads}

Production LLM inference is bursty in both compute and KV-cache usage.
To meet strict latency SLAs, online services reserve peak headroom, leaving GPUs underutilized on average.

\begin{figure}[t]
    \centering
    \includegraphics[width=0.5\linewidth]{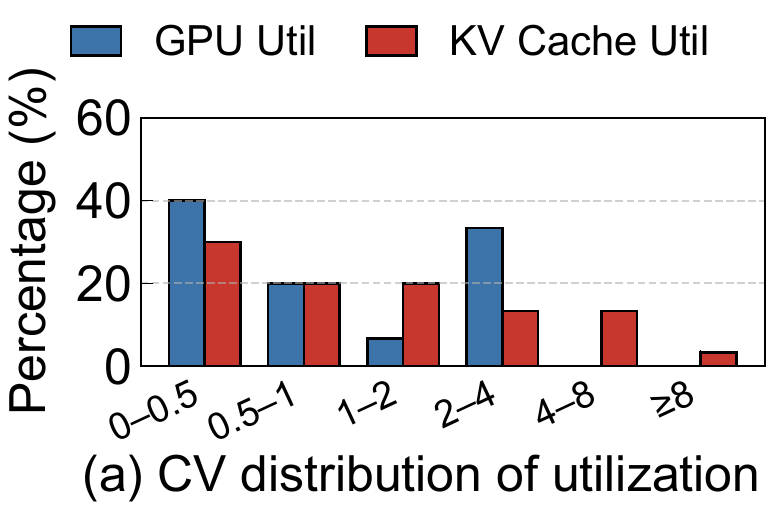}
    \caption{Coefficient of Variation (CV) analysis of GPU utilization and KV cache utilization.}
    \label{fig:eval-kvcache-cv}
\end{figure}

\begin{figure}[t]
    \centering
    \includegraphics[width=1.0\linewidth]{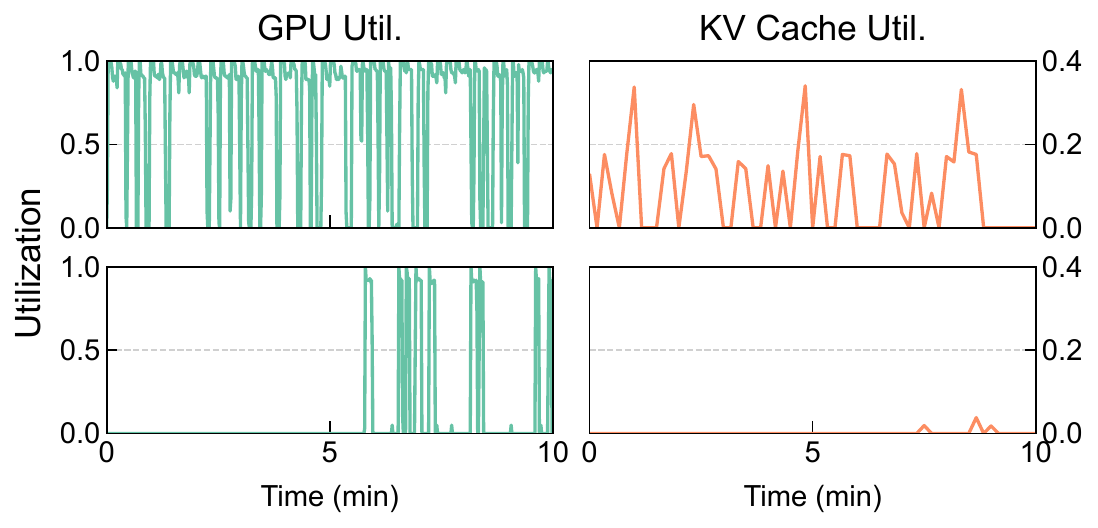}
    \caption{GPU utilization and KV cache utilization time series for user-facing inference (bottom)
    and reward model inference (top). They show different degrees of burstiness in KV cache utilization.}
    \label{fig:bg:burst-analysis}
\end{figure}

\parabf{Burstiness in compute and KV-cache memory.}
Each request uses GPU compute and allocates KV cache throughout its lifecycle.
As a result, compute utilization often switches between idle and fully busy.
Meanwhile, KV cache grows with the number of concurrent context tokens, and can spike under batch arrivals.
Figure~\ref{fig:eval-kvcache-cv} measures burstiness across workloads using CV
and Figure~\ref{fig:bg:burst-analysis} shows two typical patterns:
some workloads are bursty in both compute and KV cache,
while other workloads are bursty mostly only in compute.

\parabf{Production workloads and their SLAs.}
Production inference includes both online and offline workloads.
Online inference is user-facing (or latency-critical stages in post-training) and must meet strict latency SLAs,
so it can tolerate almost no interference.
Offline inference often requires only throughput SLAs, or has no SLA.
To meet online SLAs under bursty demand, operators overprovision GPUs, which leads to low average utilization.

% \subsection{Low GPU Utilization of Production Workloads}

% \parabf{Burstiness of LLM inference.}
% The burstiness nature of LLM inference is due to both compute and memory usage.
% When an inference request arrives, it occupies compute resources and allocates GPU memory for \texttt{KV cache},
% during autoregressive generation from prefill to decode phases.
% \textit{Computation bursts} occur when the inference system switches between idle and busy states.
% GPU compute utilization hits 100\% with any request in flight, dropping to 0\% in idle states.
% The \textit{memory usage bursts} when many requests arrive simultaneously,
% since memory usage scales with the number of context tokens being processed.
% Figure~\ref{fig:eval-kvcache-cv} shows the Coefficient of Variation (CV)
% analysis of GPU utilization and KV cache utilization in our production cluster.
% Some workloads exhibit high burstiness in both compute and memory usage (e.g. reward model inference, top of Figure~\ref{fig:bg:burst-analysis}),
% while others show high burstiness only in compute (e.g. user-facing inference, bottom of Figure~\ref{fig:bg:burst-analysis}).
% Insufficient reservation to absorb bursts causes latency spikes and SLA violations.
% Thus, operators usually overprovision GPUs, leading to low utilization.

\begin{table*}[t]
\centering
% \small
\caption{Comparison of schemes for online-offline colocation.}
\label{tab:bg:existing-schemes}
\begin{tabular}{lccccccc}
\toprule
& \textbf{TGS} & \textbf{Conserve} & \textbf{Gpreempt} & \textbf{Valve} \\
\midrule
\textbf{Compute Interference} & \textcolor{red!80!black}{\textbf{Iteration-level}} & \textcolor{red!80!black}{\textbf{Layer-level}} & \textcolor{red!80!black}{\textbf{Frequent}} & \textcolor{green!80!black}{\textbf{$<1$ms per-request}} \\
\textbf{Memory Interference} & \textcolor{red!80!black}{\textbf{Frequent}}  & \textcolor{red!80!black}{\textbf{Layer-level}} & \textcolor{red!80!black}{\textbf{Not handling}} & \textcolor{green!80!black}{\textbf{Sub-layer, limited rate}} \\
\textbf{Framework Modifications} & \textcolor{green!80!black}{\textbf{0 LOC}} & \textcolor{red!80!black}{\textbf{$>$5000 LOC}} & \textcolor{green!80!black}{\textbf{0 LOC}} & \textcolor{green!80!black}{\textbf{$<$20 LOC}} \\
\textbf{Driver Modifications} & \textcolor{green!80!black}{\textbf{0 LOC}} & \textcolor{green!80!black}{\textbf{0 LOC}} & \textcolor{red!80!black}{\textbf{$>$200 LOC}} & \textcolor{green!80!black}{\textbf{1 LOC}} \\
% \textbf{Nearline SLA guarantee} & \textcolor{red!80!black}{\textbf{NO}} & \textcolor{green!80!black}{\textbf{YES}} & \textcolor{red!80!black}{\textbf{NO}} & \textcolor{red!80!black}{\textbf{NO}} & \textcolor{red!80!black}{\textbf{NO}} \\
\bottomrule
\end{tabular}
\end{table*}

\subsection{Key Requirements for Inference Online-Offline Colocation in Production}

\parabf{Extremely low interference for online workloads.}
To meet strict SLAs for real-time online inference, the system must introduce almost no extra delay.
Interference comes from both compute and memory effects, and is driven by \emph{how long} each preemption lasts and \emph{how often} preemptions happen. Thus, the key goal is to bound both
\emph{preemption latency} and \emph{preemption rate}.

\parabf{Minimal modifications to drivers and frameworks.}
Production deployment requires minimal modifications to GPU drivers and inference frameworks.
Extensive modifications increase maintenance burden and limit broader adoption.

\parabf{Throughput SLAs for offline workloads.}
The system should place offline workloads on suitable nodes to meet throughput SLAs, while maximizing cluster throughput.

\subsection{Existing Solutions for Online-Offline Colocation}

A complementary line of work colocates latency-critical online inference with offline jobs on the same GPU,
to backfill idle capacity while preserving strict online SLAs~\cite{wu2023transparent,han2022microsecond,
qiao2024conserve,fan2025gpreempt,shen2025xsched,prabhu2025vattention,
xu2024vtensor,yu2025prism}. Table~\ref{tab:bg:existing-schemes} compares these systems.
In production, they commonly face two obstacles: (i) noticeable interference with online workloads
(due to slow or frequent preemptions), and
(ii) extensive driver/framework modifications that make deployment and maintenance hard.

\begin{figure}[t]
    \centering
    \includegraphics[width=0.6\linewidth]{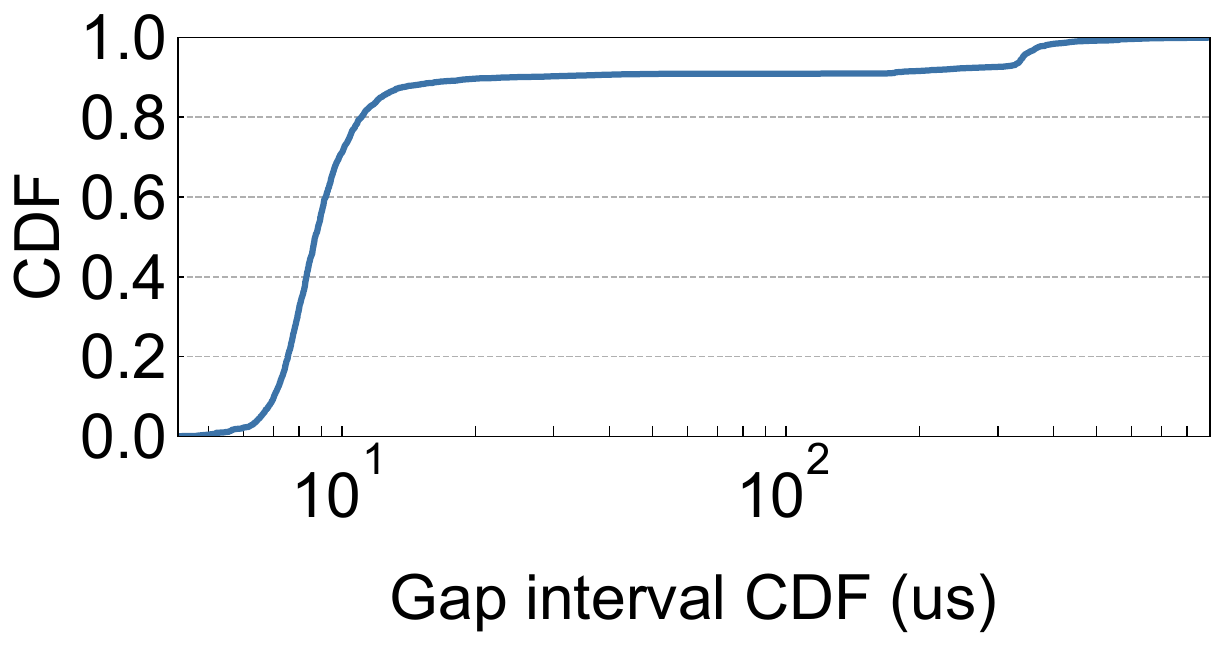}
    \caption{Distribution of gap interval between decode iterations.}
    \label{fig:gap}
\end{figure}

\parabf{Compute-side preemption.}
TGS~\cite{wu2023transparent}, XSched~\cite{shen2025xsched} Lv2 intercepts kernel launches, but LLM serving often relies on CUDA Graphs,
which bundle kernels of one inference iteration into a single graph,
so preemption degrades to graph-level granularity.
Conserve~\cite{qiao2024conserve} inserts checkpoints into inference code and preempts at the transformer-layer level,
but long prefills in production batch inference (e.g., 32k tokens) can stretch layer-level preemption delay to hundres of milliseconds.
The preempted iteration is duplicated, decreasing offline throughput.
Gpreempt~\cite{fan2025gpreempt} uses a CUDA-driver timeslice for automatic switching, but the decode phase has short gaps between iterations (Figure~\ref{fig:gap}),
so it may run offline kernels once per iteration, causing frequent preemptions and increased queue length.

% Online-offline colocation aims to backfill idle capacity with offline work
% and allows preemption of offline jobs during online bursts.
% Existing approaches feature coarse-granularized or frequent preemptions in both compute and memory,
% leading to significant interference on online workloads.
% Specifically, TGS~{wu2023transparent} preempts compute in iteration-level,
% ConServe~\cite{qiao2024conserve} preempts compute in layer-level and memory in iteration-level.
% Coarse preemption granularity causes extremely long delays when online jobs arrive during offline prefill phase,
% increasing online tail TTFT to ~100x of nominal.
% Gpreempt~\cite{fan2025gpreempt} achieves microsecond-scale compute preemption by setting timeslice for workloads.
% but it incurs a high frequency of preemptions during the decoding stage.
% The uncontrollable GPU timeslice mechanism unaware of LLM inference patterns
% wakes up the offline workload once between each decode iteration,
% leading to increased TPOT of online workloads.
% Besides, many of the above rely on inference framework modifications or hardware driver
% modifications, limiting transparency and production adoption.
% Consequently, none of them are able to achieve broad production adoption.

\parabf{Memory-side KV-cache isolation.}
vAttention, Conserve, vTensor, and Prism virtualize KV-cache placement via VMM indirection and resize memory footprints on demand~\cite{prabhu2025vattention,qiao2024conserve,xu2024vtensor,yu2025prism}.
However, they do not fully solve how to reclaim offline KV memory quickly and safely when online demand spikes.
For example, Conserve~\cite{qiao2024conserve} can reclaim KV cache only at transformer-layer boundaries, which can delay reclamation by up to hundreds of milliseconds during long prefills.
Moreover, prior work largely does not discuss how to control the reclamation frequency.
In practice, inference workloads change their memory regions and sizes over time;
naively relying on UVM~\cite{cuda_unified_memory} or aggressively sharing memory can trigger reclamation repeatedly, causing severe interference to online workloads.

% When the KV cache usage of online workloads bursts, we have to wait until the
% finish of the present offline iteration to evict offline page by either swapping out
% or discarding KV cache.
% TODO: rethink REEF and XSched
\parabf{Deployability.}
Many systems also fall short in deployability.
Conserve~\cite{qiao2024conserve} requires extensive inference-framework changes,
often involving thousands of lines of code (e.g., injecting checkpoints into inference code and adding new scheduling modules).
REEF~\cite{han2022microsecond} requires replacing the compiler toolchain with a custom compiler, which in turn demands major changes to the user container image. XSched~\cite{shen2025xsched} Lv3 and REEF only support idempotent operators, but all-reduce in multi-GPU inference and some linear attention kernels are not idempotent,
making the whole CUDA graph unpreemptible.
As a result, no existing approach simultaneously achieves low-interference, LLM-compatible preemption and production-friendly deployability.

\subsection{Challenges}

We face three challenges in designing a production-friendly online-offline inference colocation system in production.

\parabf{Challenge 1: Gap-aware sub-millisecond compute interference.}
The system must achieve sub-millisecond compute preemption
while not inserting offline wake-ups between online decode iterations.
This requires controllable, fast offline switch, with awareness of online request lifecycles.

\parabf{Challenge 2: Rate-limited sub-layer memory reclamation.}
Online bursts may require reclaiming offline KV cache \emph{immediately} in sub-layer granualrity.
However, swapping KV pages to CPU is too slow, while invalidating KV pages without coordination
can cause illegal accesses. The system must reclaim memory promptly while not killing offline applications.
Restricting the reclamation frequency with framework transparency is also critical.

\parabf{Challenge 3: Precise offline performance modeling.}
Offline throughput varies with online burstiness and multi-GPU behaviors,
requiring precise modeling and scheduling.

\section{\sysname Overview}
\label{sec:overview}

\begin{figure}[t]
    \centering
    \includegraphics[width=1.0\linewidth]{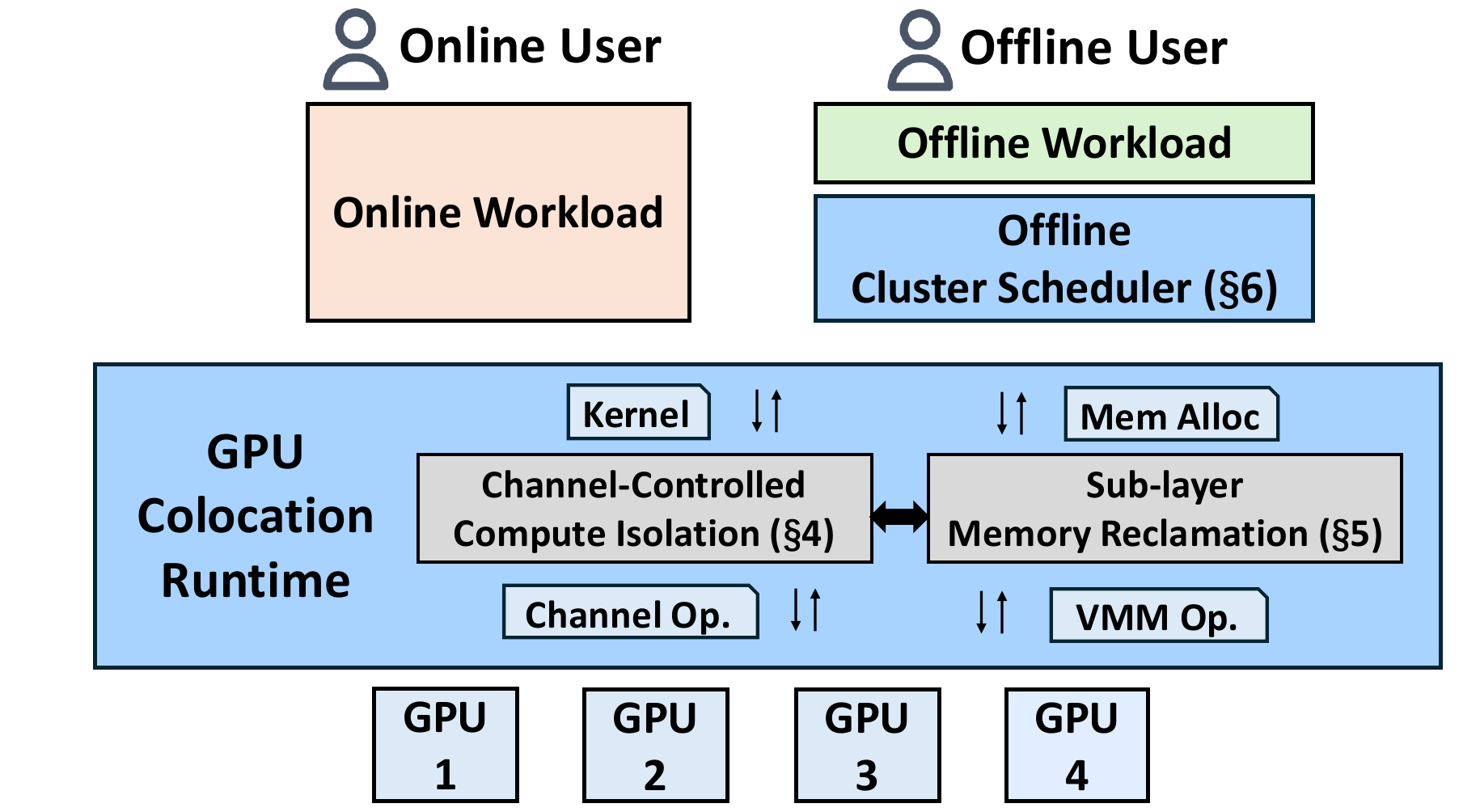}
    % \vspace{-5mm}
    \caption{\sysname architecture overview.}
    % \vspace{-4mm}
    \label{fig:design:cluster}
\end{figure}

We introduce \sysname, an industrial system for online-offline inference colocation.
\sysname is designed to meet three goals: (1) low compute and memory interference to online
workloads, (2) reliable throughput SLAs for offline workloads, and
(3) minimal framework/driver modifications. 
Figure~\ref{fig:design:cluster} shows the overall architecture.

At the node level, the \emph{GPU Colocation Runtime}
enables compute and memory sharing with low interference
by jointly bounding preemption latency and rate.
For compute, it limits online impact by providing
sub-millisecond, infrequent kernel preemptions via channel control
and workload-aware offline execution control (\S\ref{sec:compute}).
For memory, it follows prior work~\cite{prabhu2025vattention} to
share GPU memory through a global pool with coarse-grained handles
and an \emph{allocate--release} interface. It bounds memory interference with
fast sub-layer memory reclamation coordinated with compute preemption,
and controls reclamation frequency via MIAD(Multiple Increase, Addictive Decrease)-style online reservation
(\S\ref{sec:memory}).
These mechanisms also preserve high offline throughput by harvesting most idle compute cycles; during memory reclamation, selective eviction affects fewer offline requests and safe sub-layer reclamation avoids terminating offline workloads.

These mechanisms also preserve
high offline throughput by harvesting most idles compute cycles, impacting less offline requests during memory preemptions, without terminating offline workloads during memory preemptions.

At the cluster level, online workloads are submitted directly to GPUs,
while offline workloads are submitted to the \emph{Cluster Scheduler}
(\S\ref{sec:scheduler}). The scheduler
builds a comprehensive performance model of offline workloads on harvested GPUs,
and schedules them to satisfy their throughput SLAs, which is
specified as the fraction of the standalone throughput. 
\section{Channel-Controlled Compute Isolation}
\label{sec:compute}

We use workload-aware \emph{channel control} to achieve two goals:
\textbf{sub-millisecond} preemption latency and \textbf{at most one} preemption per online request.
Channel control provides a fast and precise way to pause and resume offline execution,
while workload awareness triggers preemption in time and avoids frequent preemptions.

\begin{figure}[t]
    \centering
    \includegraphics[width=\linewidth]{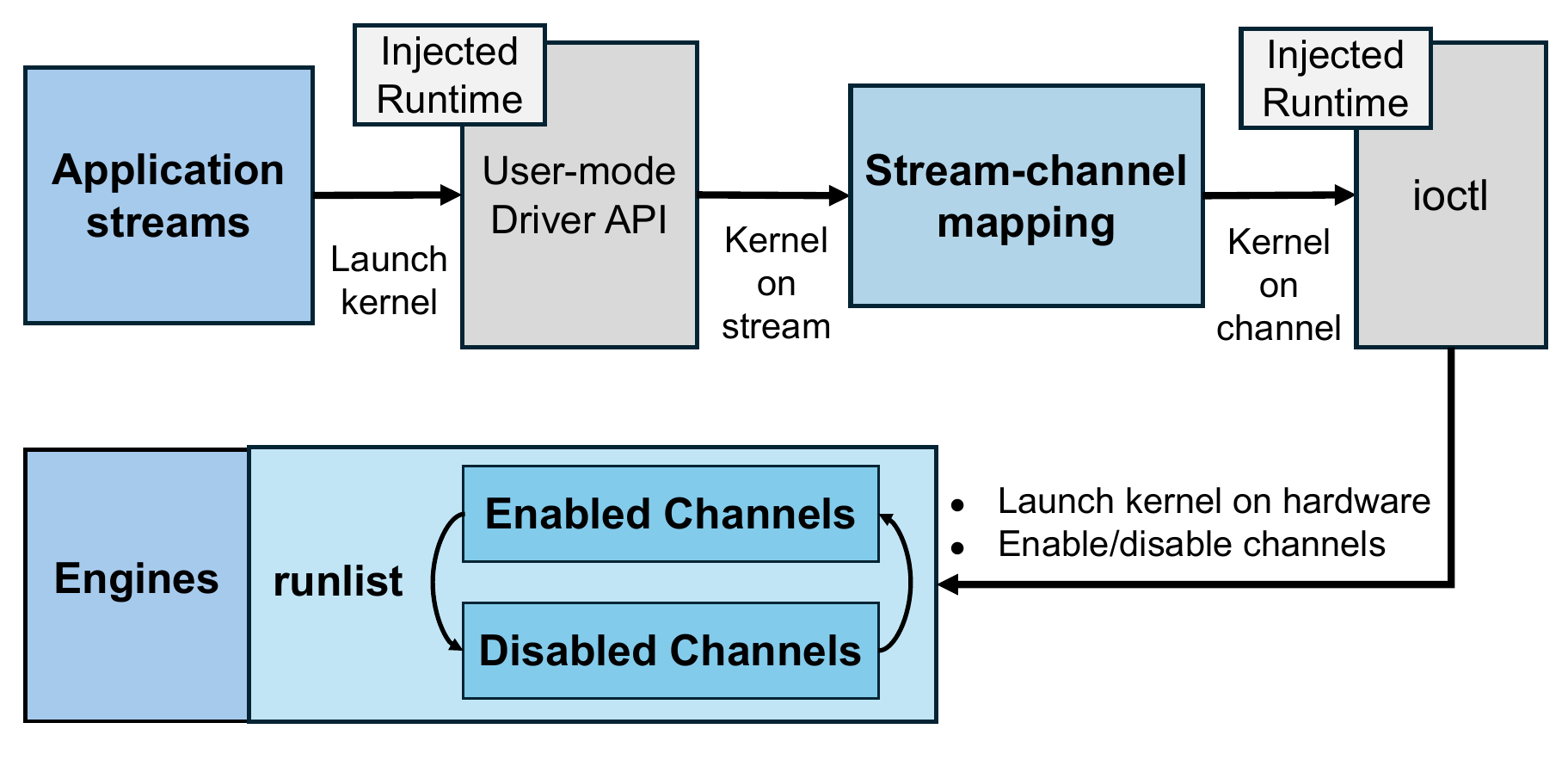}
    \caption{Channel control in the kernel launch path.}
    \label{fig:channel}
  \end{figure}

\subsection{GPU Channel Control}
\label{sec:compute:channel}

\paraf{Channels in the kernel launch path.}
Figure~\ref{fig:channel} shows the kernel launch path and where GPU channels fit in.
A CUDA stream issues kernel launches through the user-mode driver,
which submits work to a \emph{channel} managed by the kernel-mode driver (KMD) and the GPU.
A process typically owns one or more channels, which are bound to specific compute engines.
The GPU schedules channels using a hardware-maintained \emph{runlist}.
Importantly, the KMD exposes standard ioctl interfaces to create, enable, and disable channels,
and to submit work to channels.
These ioctls (I/O control commands) make channels a practical control point for controlling offline compute.
% These ioctls form the control surface we later leverage: by identifying the
% channel handles used by a workload, we can disable/enable the channel
% to pause and resume already-launched kernels.

\parabf{Channel control for portable compute isolation.}
\sysname uses this ioctl-level control point to preempt and resume offline workloads with low latency.
Specifically, it disables the offline workload's channel to preempt execution and later re-enables it to resume.
On Pascal+ GPUs, disabling a channel triggers a hardware context save to an on-GPU context-save buffer,
so in-flight kernels can be safely restored after re-enable (e.g., registers and other on-chip state).
These operations take effect within 1 ms, enabling fast preemption without waiting for kernel boundaries.

A practical challenge is that these KMD ioctls require driver-managed identifiers (e.g., GPU-client and channel handles)
that are not exposed through CUDA APIs.
We obtain these identifiers without modifying the driver by intercepting CUDA initialization ioctls:
their arguments contain the GPU-client and channel identifiers, which consistently appear as matched pairs on Pascal+ drivers.
Our colocation runtime records a mapping from each application to its (GPU-client, channel) handles, and then issues the
corresponding disable/enable ioctls on demand.

\parabf{Optimizing preemption latency for multi‑GPU preemption.}
Naively issuing preemption ioctls on a multi-GPU node leads to latency that grows roughly linearly with the number of GPUs.
The bottleneck is a shared write lock that the KMD holds while handling these ioctls across GPUs on the same node.
We find this synchronization is not required for inference tasks.
On Turing+ GPUs, the preemption command can be offloaded directly to the target device without taking the global lock,
reducing kernel-space overhead.
Accordingly, we apply a one-line driver modification which changes the flag used to bypass the lock and offload the command to the channel. 
With this modification, preemption latency on an 8-GPU node drops from $>5$\,ms to $<1$\,ms.

\begin{figure}[t]
  \centering
  \includegraphics[width=\linewidth]{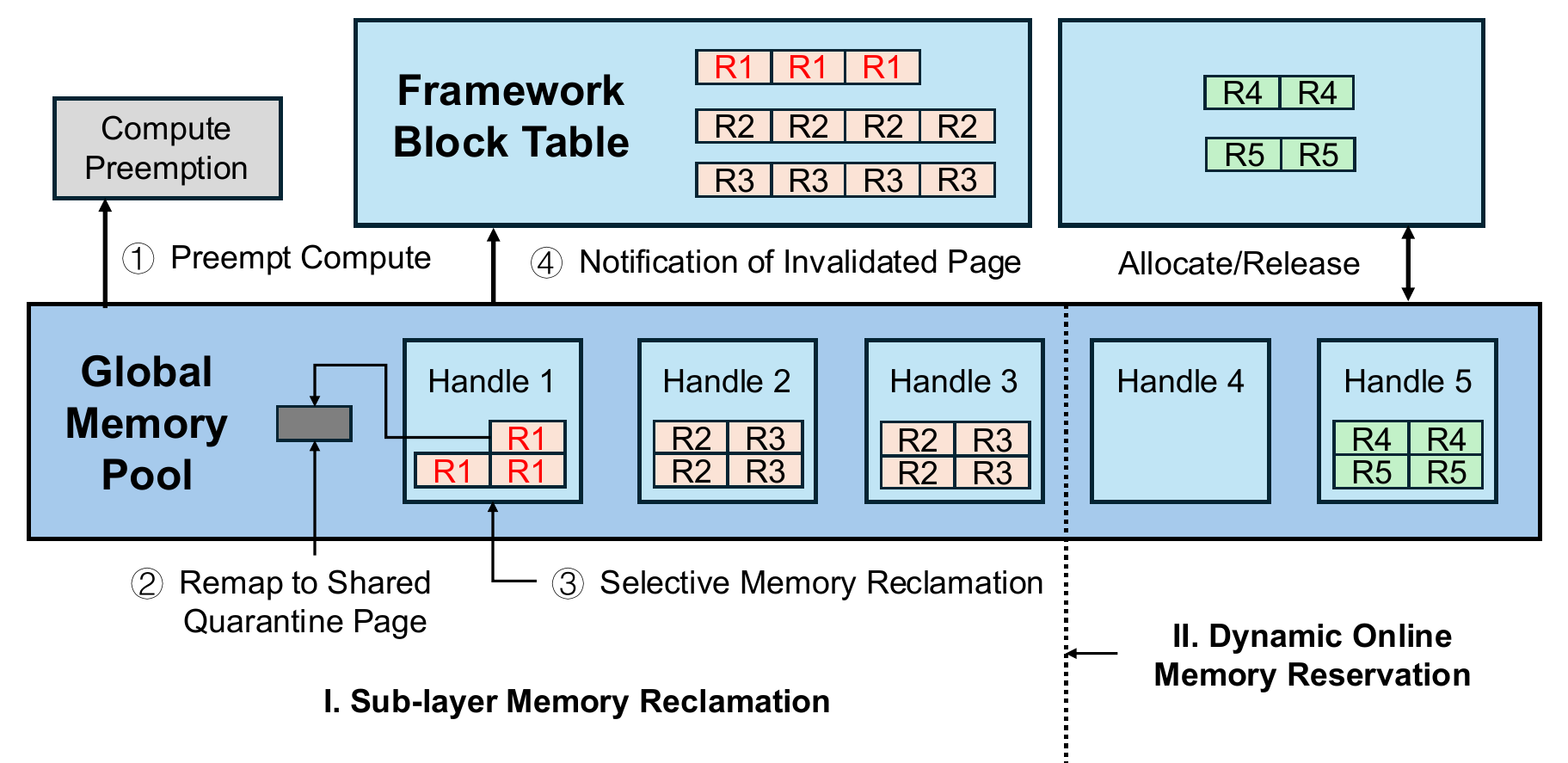}
  \caption{Sub-layer memory reclamation with dynamic online memory reservation.}
  \label{fig:memory-preempt}
\end{figure}

\subsection{On-GPU Offline Workload Scheduling}
\label{sec:compute:schedule}

\parabf{Preempting offline workloads.}
Our runtime, injected into the online process, intercepts kernel-launch commands to track whether the online workload is active.
When the online workload transitions to busy, the runtime immediately issues channel-disable commands for all offline workloads on the node,
so offline execution is paused promptly.

\parabf{Waking up offline workloads.}
To bound preemption to at most once per online inference request, we do not re-enable offline workloads immediately when the online workload becomes idle,
since short idle gaps can appear between decode iterations.
Instead, we wake up offline workloads only after a cooldown interval $T_{\text{cool}}$ during which the online workload stays continuously idle.
We set $T_{\text{cool}}$ to twice the maximum gap $G$ between decode iterations, as measured by our runtime instrumentation.
This avoids waking offline work in per-iteration gaps and ensures at most one preemption over the lifetime of an online request.
\section{Sub-layer Memory Reclamation with Dynamic Reservation}
\label{sec:memory}

% Mapping KV cache to virtual memory by CUDA VMM API provides the opportunity
% to transparently harvest the unused memory of online workloads to run offline
% workloads. Unfortunately, for online workloads, these solutions lead to
% memory interference when online memory usage bursts. Online workloads need to
% wait for offline KV cache eviction, which requires either GPU memory swapping
% out, or pending until iteration finish. Besides, the remap of virtual memory
% of KV cache to physical memory also incurs extra mapping overheads.
% To eliminate these interferences, we propose \textit{sub-layer-level
% memory preemption} and \textit{dynamic online memory reservation} to largely
% eliminate these interferences.
The main bottleneck of online-offline memory sharing is KV-cache reclamation.
When online memory demand spikes, the system may need to unmap and remap offline VMM-mapped KV cache~\cite{prabhu2025vattention}
on the online critical path.
\sysname addresses this with three techniques:
(i)~\emph{sub-layer memory reclamation} to make preemption fast,
(ii)~\emph{dynamic online memory reservation} to reduce interference rate,
and (iii)~\emph{selective handle reclamation} for higher offline throughput without
adding interference.

\parabf{Sub-layer memory reclamation.}
We first focus on reducing reclamation latency.
A natural baseline~\cite{xiang2025aegaeon,qiao2024conserve} reclaims KV memory only at iteration boundaries or layer boundaries, because unmapping KV pages during kernel execution can lead to unrecoverable memory faults.
However, this coarse granularity can make reclamation slow,
especially when online bursts arrive during a long prefill iteration.
prefill phase
when online bursts arrive during offline prefill iterations.

\sysname enables safe sub-layer reclamation by coordinating compute and
memory preemption (Figure~\ref{fig:memory-preempt}). When online workloads need memory,
we always disable offline compute first, ensuring
no in-flight kernel can access pages being reclaimed. We then select a set of offline ``evictor'' memory handles, remap the virtual pages in those handles to a shared
quarantine page, and finally reclaim the memory handles for online workloads. This avoids faults and allows the offline workload to resume later.

Accesses to reclaimed pages can lead to incorrect intermediate tokens. To keep the behavior correct,
\sysname records the reclaimed KV page (block) IDs and exposes them through a small patch:
a single callback that returns the invalidated IDs for each request after a decode step.
Across vLLM, SGLang, and TensorRT-LLM, this integration touches at most two scheduler-side functions
and requires fewer than 20 lines of code changes.
The framework then discards intermediate data for affected requests, returns them to the waiting state with only the input and previously generated tokens, and later resume them by recomputing.
This achieves sub-layer reclamation without crashes.

\parabf{Dynamic MIAD-style memory reservation.}
To reduce reclamation frequency within a given budget while maximizing offline memory,
\sysname maintains a dynamic online KV-cache headroom $H$ as pre-mapped VMM handles.
\sysname adapts $H$ using MIAD (Multiplicative Increase, Additive Decrease):
on an online pressure event (i.e., when $H$ reaches 90\% utilization), it multiplicatively increases $H$ by a factor $\alpha$
to reserve more mapped handles in advance; when pressure is absent, it shrinks conservatively by releasing one handle every interval $T$.
The release interval $T$ is also MIAD-controlled: if the pressure-event rate over a sliding window exceeds the user-specified target,
\sysname multiplicatively increases $T$; otherwise, it decreases $T$. This reservation drives the reclamation rate toward the target.

% \begin{figure}[t]
%   \centering
%   \includegraphics[width=\linewidth]
%     {figures/design/memory/miad_mechanism.pdf}
%   \caption{MIAD for reserving memory for online workloads.}
%   \label{fig:miad}
% \end{figure}

\parabf{Selective handle reclamation.}
The KV cache is not allocated continuously on the memory handle due to the memory fragment problem,
so one handle can be shared by different numbers of offline requests.
Reclaiming a handle blindly may preempt more requests than necessary.
\sysname uses \emph{selective} handle reclamation (Algorithm~\ref{alg:selective-reclaim}) to minimize the number of affected offline requests.
Specifically, it greedily selects handles with the lowest marginal token cost, defined as the total number of extra tokens incurred by the additional requests affected by reclaiming that handle.
This improves offline throughput without increasing online interference.

\begin{algorithm}[t]
\caption{\sysname selective memory reclamation}
\label{alg:selective-reclaim}
\begin{algorithmic}[1]
\REQUIRE Number of handles $k$; handle set $\mathcal{H}$ (equal size);
request cost $\textsc{Cost}(r)$; impacted requests $\textsc{Reqs}(h)$
\ENSURE A handle subset $\mathcal{S}$ to reclaim
\STATE $\mathcal{S}\leftarrow \emptyset$, $\mathcal{E}\leftarrow \emptyset$
\FOR{$i=1$ to $k$}
  \STATE $h^{*}\leftarrow \arg\min_{h\in\mathcal{H}\setminus\mathcal{S}}
  \sum_{r\in \textsc{Reqs}(h)\setminus\mathcal{E}} \textsc{Cost}(r)$
  \STATE $\mathcal{S}\leftarrow \mathcal{S}\cup\{h^{*}\}$
  \STATE $\mathcal{E}\leftarrow \mathcal{E}\cup \textsc{Reqs}(h^{*})$
\ENDFOR
\STATE \textbf{return} $\mathcal{S}$
\end{algorithmic}
\end{algorithm}

\section{\sysname Cluster Scheduling}
\label{sec:scheduler}

% To satisfy offline throughput SLAs while improving overall utilization, we
% comprehensively model how resource average, resource burstiness, and
% multi-card behavior affect offline LLM inference performance on harvested
% GPUs. Building on this model, our scheduler reasonably matches tasks to
% resources to maximize throughput without compromising throughput SLAs.

To satisfy offline throughput SLAs, we build a comprehensive performance model
for offline LLM inference on harvested GPUs.
We characterize a harvested GPU along three aspects:
(i) idle compute fraction; (ii) the burstiness and average
of memory usage; and (iii) the multi-GPU behavior of online workloads.
We formulate it as
\begin{equation}
\label{eq:thrput_model}
\frac{\operatorname{Thrput}_{(w,N)}}{\operatorname{Thrput}_{(w,\text{max})}}
= P_{\text{compute},(w,N)}\cdot P_{\text{memory},(w,N)}\cdot P_{\text{multi},(w,N)}.
\end{equation}
Here, $w$ denotes an offline workload and $N$ denotes a node.
$\operatorname{Thrput}_{(w,N)}$ is the effective throughput of $w$ on $N$, and
$\operatorname{Thrput}_{(w,\text{max})}$ is its throughput on the monopolized GPU.
We define the three performance factors as follows.

\parabf{Idle compute fraction.}
We measure the idle compute fraction using the colocation runtime as the
fraction of GPU timeslices available to run the offline workload.

\parabf{Burstiness and average of memory.}
GPU memory determines both feasibility and throughput. For each workload $w$, we profile it once at submission to obtain a memory--throughput curve $\operatorname{Thrput}_w(\text{memory})$.
Let $M$ be the available memory on node $N$. Without eviction, the effective throughput is the time average of $\operatorname{Thrput}_w(M)$ over the node's memory trace.
When $M$ dips below the workload's required memory $M_{\text{req}}$, the shirnk $\Delta M=\max(0, M_{\text{req}}-M)$ introduces throughput loss.
We use a workload-specific coefficient $\operatorname{MAC}_w$ to map the expected deficit to throughput loss. The memory factor is formulated as:
\begin{equation}
\label{eq:memory_factor}
P_{\text{memory},(w,N)} =
\frac{\mathbb{E}\!\left[\operatorname{Thrput}_w(M)\right]
- \operatorname{MAC}_w\cdot \mathbb{E}\!\left[\Delta M\right]}
{\operatorname{Thrput}_w(M_{\text{max}})}.
\end{equation}

\parabf{The multi-card behavior.}
Online multi-GPU services often use GPUs asynchronously, so activity can be misaligned across cards; in our trace, 32\% of instances show only partial overlap.
In contrast, model-parallel offline inference runs in lockstep. Misalignment across cards then creates stragglers and idle gaps, reducing throughput and risking SLA violations.
We quantify cross-card alignment with a pairwise score
$P_{\text{multi}, (w,N)}=\frac{T_{\cap}(N)}{T_{\cup}(N)}$, where $T_{\cap}$ is overlapping busy time and $T_{\cup}$ is union busy time.
At placement, we admit a $k$-GPU job only if all pairs satisfy $P_{\text{multi}, (w,N)}\ge 0.95$.

% \subsection{Offline Workload Scheduling}
% \label{sec:scheduler:workload}
\parabf{Scheduling.}
Building on this model, our scheduler schedules offline workloads to
nodes with guaranteed throughput SLA. Besides, a monitor periodically checks
the past throughput of each offline throughput and evicts those that
persistently violate their SLA for rescheduling.

% We first classify each node by its average available memory and idle compute
% percentage. For a workload $w$, we use these averages to identify feasible
% node classes and sample a small set of candidate nodes (for multi-GPU jobs,
% we sample nodes that can provide $k$ cards). We then run burstiness-aware
% prediction on the sampled candidates, filter out those whose predicted
% throughput would violate the SLA, and select the final node and card set.
% Finally, to handle long-term SLA violations, the runtime monitors throughput
% via the page allocation rate; if a workload persistently falls below its SLA,
% we evict and reschedule it.

\section{Evaluation}
\label{sec:evaluation}

\begin{figure}[t]
    \centering
    \includegraphics[width=0.36\textwidth]{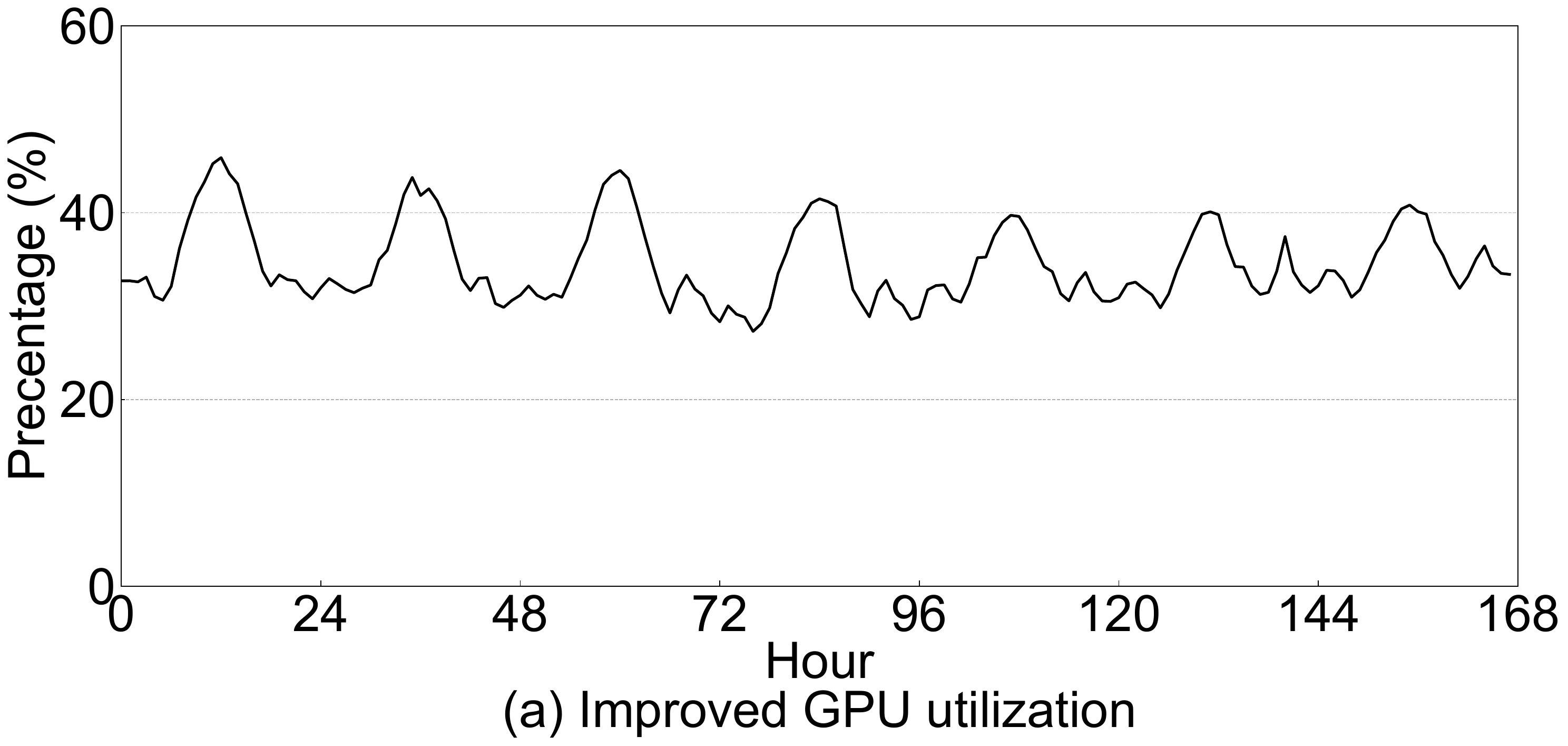}
    \caption{Cluster GPU utilization with \sysname.}
    \label{fig:eval1a}
\end{figure}

\begin{figure}[t]
    \centering
    \includegraphics[width=0.36\textwidth]{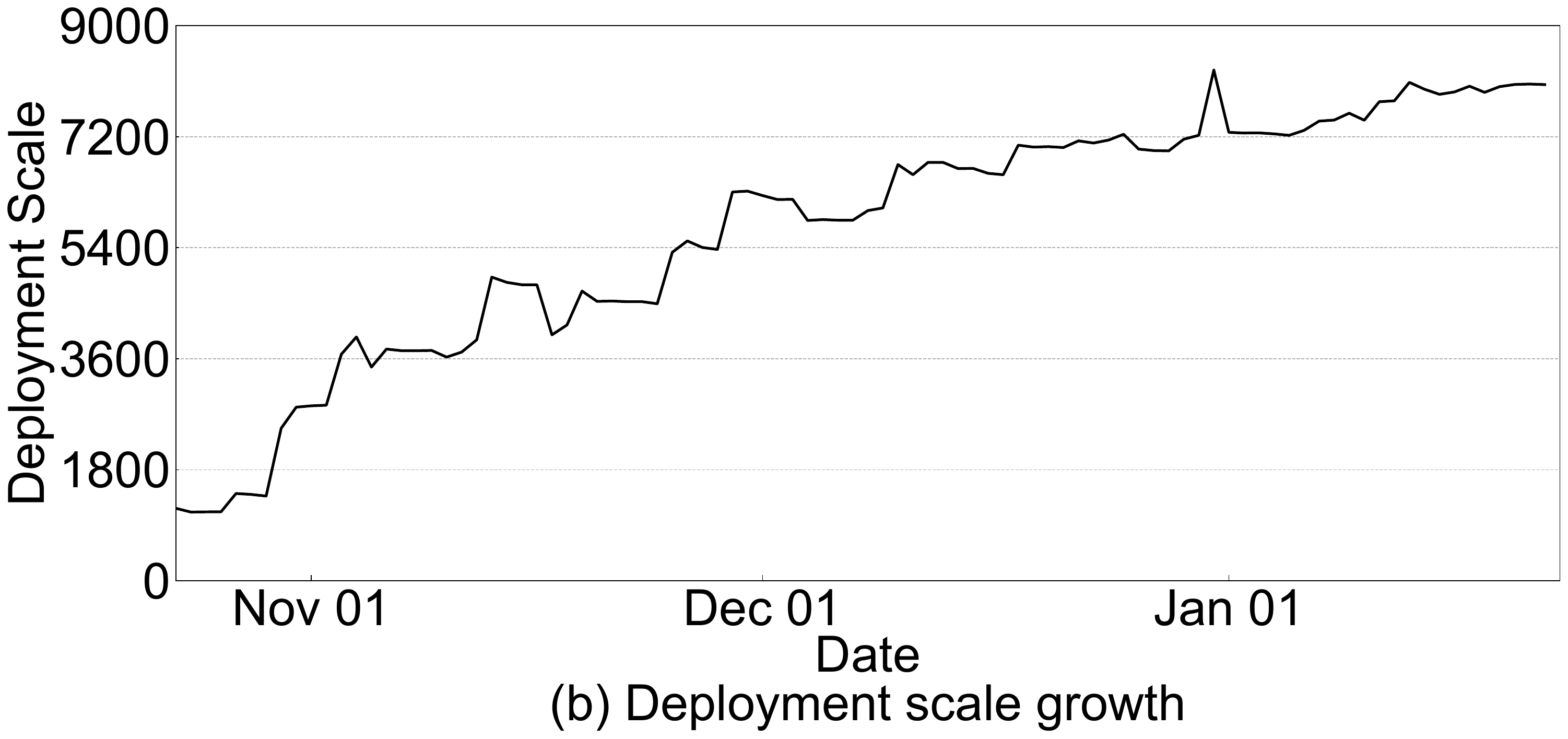}
    \caption{Deployment scale with \sysname.}
    \label{fig:eval1b}
\end{figure}

In this section, we first quantify the production impact of \sysname
(\S~\ref{sec:eval:utilization}), then compare it with existing online-offline colocation approaches
in terms of interference to online workloads and throughput of offline workloads (\S~\ref{sec:eval:comp}).
% and finally evaluate the effectiveness of the cluster scheduler (\S~\ref{sec:eval:scheduler}) on higher offline throughput.

\subsection{Production Impact of \sysname}
\label{sec:eval:utilization}

\paraf{Deployment.}
\sysname is deployed in production clusters with 8045 GPU cards.
The clusters serve both online and offline inference workloads.
\sysname has been in production for more than three months, with the number of managed
GPU cards increasing over time, illustrated by Figure~\ref{fig:eval1b}.

\paraf{Metrics.}
We use two metrics to quantify \sysname's end-to-end impact: (i) \emph{
improved GPU utilization}, which is the fraction of time when GPUs execute offline compute;
and (ii) \emph{saved GPU cards}. The amount of GPU cards saved by each colocated offline workload
is computed as the throughput normalized by standalone offline throughput.

\parabf{Results.}
Figure~\ref{fig:eval1a} shows improved GPU utilization in the production cluster
over one week, with an average of 34.6\% improvement.
The inference work done by the offline workloads translates to a saving of 2170 GPU cards.

\subsection{Comparison with Alternative Colocation Approaches}
\label{sec:eval:comp}

\paraf{Metrics.}
We evaluate both the interference to online workloads and the throughput of offline workloads using three metrics:
\begin{itemize}
    \item TTFT increase percentage: The increase percentage of online workloads under each strategy, compared with standalone running TTFT.
    \item TPOT increase percentage: The increase percentage of online workloads under each strategy, compared with standalone running TPOT.
    \item Offline throughput: To compare the throughput of different strategies clearly, we report the normalized throughput, which is the throughput divided by the throughput under no-memory-preemption, i.e. Prism with our compute preemption (Channel+Prism).
\end{itemize}
% The interference is measured by the increase percentage of average online TTFT and TPOT.
% The throughput is measured by the colocation throughput.

\begin{figure}[t]
    \centering
    \includegraphics[width=0.5\textwidth]{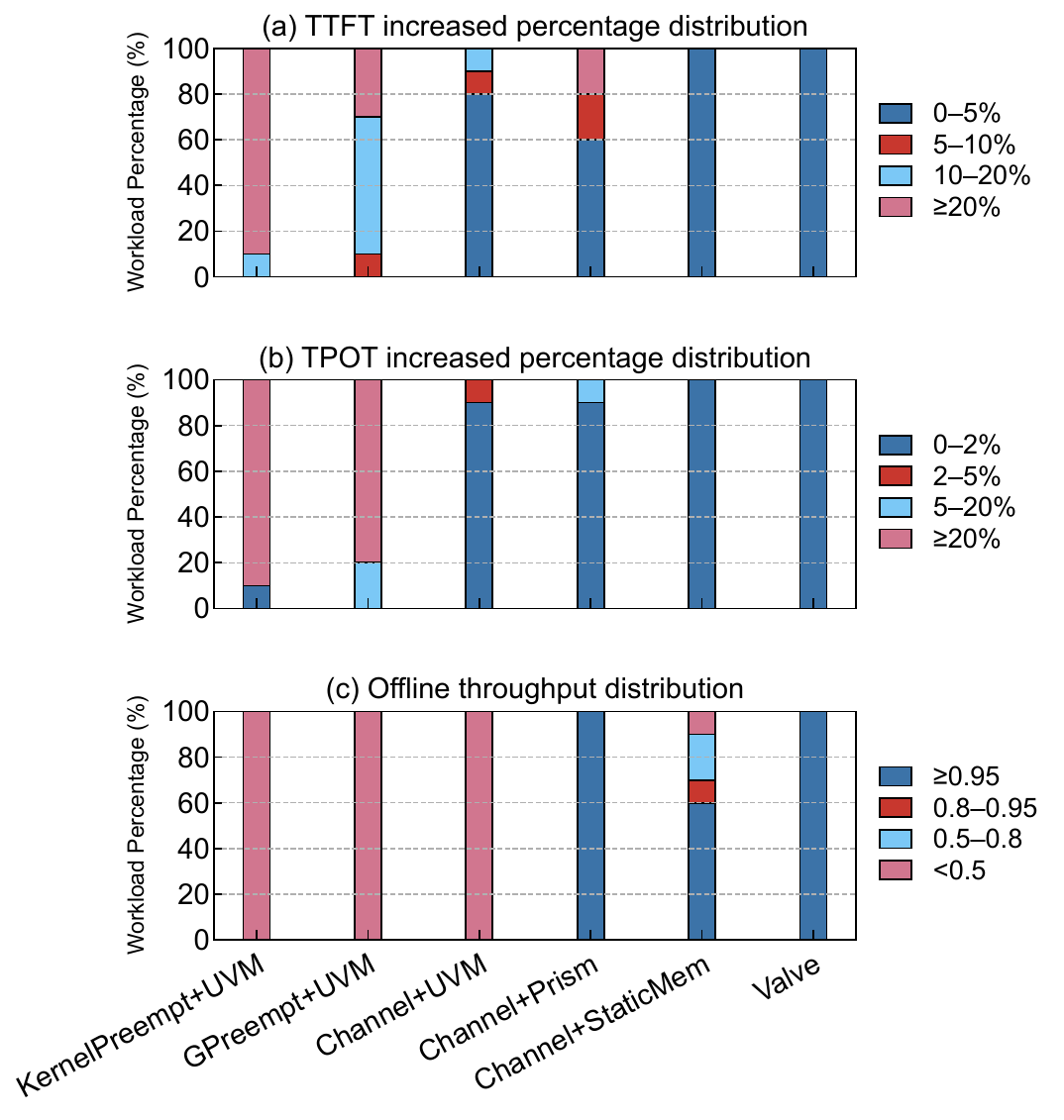}
    \caption{Distribution over 10 workloads of (a) TTFT increase percentage (b) TPOT increase percentage and (c) offline throughput under each strategies. Stacked segments indicate the workload percentage in each bin.}
    \label{fig:comparison-existing}
\end{figure}

\parabf{Methodology.}
We sample 10 online workload and offline workload pairs from production deployments,
and replay them in our test cluster with different colocation strategies.

\parabf{Baseline techniques.}
We consider two orthogonal design dimensions—compute preemption and memory preemption.

\emph{Compute preemption.}
(1) KernelPreempt: kernel-level preemption—switch at kernel boundaries, adopted by TGS~\cite{wu2023transparent}.
With CUDA graphs, kernel boundaries align with iteration steps, resulting in coarse-grained preemption.
(2) GPreempt: immediate online workload preemption via setting short time slice for
offline workloads and long time slice for online workloads, proposed by GPreempt~\cite{fan2025gpreempt}.
(3) Channel: our channel-based preemption with workload awareness (\S\ref{sec:compute}).

\emph{Memory preemption.}
(1) UVM~\cite{cuda_unified_memory}: allocate normal device memory for online workloads and use CUDA Unified Virtual Memory
for offline workloads, which allows the online workloads to reclaim offline memory as needed.
% (2) IterEvict: evict at iteration boundaries only (e.g., release activations/working sets at step ends).
% (3) Kill: kill the offline workload when online workloads are short of memory.
(2) Prism~\cite{yu2025prism}: share memory between online and offline workloads via CUDA VMM.
(3) StaticMem: statically allocate unused memory to offline workloads via the CUDA VMM.
The minimum free memory over the past hour is used as the offline limit; online bursts above this
kill offline workloads immediately.
(4) OurMem: our memory preemption(\S\ref{sec:memory})—sub-layer reclamation and MIAD-style dynamic memory reservation
for low reclamation rate.

% 一些可能被问到的baseline, for rebuttal
% online offline 都用 UVM 会怎样，端到端应用 TGS 方案会怎么样？需要实验
% layer level preemption (conserve)? 很不透明

\parabf{Baseline combinations.}
We evaluate combinations of the above techniques as baselines to assess how
\sysname's preemption mechanisms reduce interference while sustaining
high offline throughput.
We (i) compare KernelPreempt+UVM, Gpreempt+UVM, and Channel+UVM to show the effects of our compute preemption;
(ii) compare Channel+UVM, Channel+Prism, Channel+StaticMem, and Valve to show the effects of our memory preemption.

\parabf{Results.}
Figure~\ref{fig:comparison-existing} illustrates the results.
\sysname\ keeps the online TTFT increase within 5\%
and TPOT increase within 2\% across all workloads, which is significantly
lower than all baselines.
Meanwhile, \sysname maintains similar offline throughput to Channel+Prism where
offline KV cache is not reclaimed, and significantly outperforms UVM-based baselines
and static memory allocation baselines.

For compute preemption,
KernelPreempt suffers from large single preemption latency in all cases
since CUDA graphs make preemption iteration-level;
Gpreempt incurs frequent preemptions because an offline wakeup and preemption happens after each inference iteration. This harms TPOT and increases queue length, which in turn increases TTFT.
In contrast, Valve's channel-based control with workload awareness
bounds both preemption latency and rate, achieving low compute interference.

For memory preemption, Channel+UVM preempts often: UVM lets offline workloads fill spare online memory,
which is reclaimed whenever online demand spikes; it also cannot use memory already allocated by online workloads,
limiting offline throughput.
Channel+Prism does not reclaim memory, forcing online batch-size reduction and more queueing, increasing TTFT significantly in 4 sampled workloads.
Channel+StaticMem shows low interference but
static allocation underutilizes memory, yielding 9\%--100\% lower offline throughput in the 4 sampled online workloads with bursty memory usage.
\sysname utilizes MIAD-style dynamic reservation and sub-layer fast reclamation
for low memory interference while keeping high offline throughput.

\begin{figure}[t]
    \centering
    \includegraphics[width=0.3\textwidth]{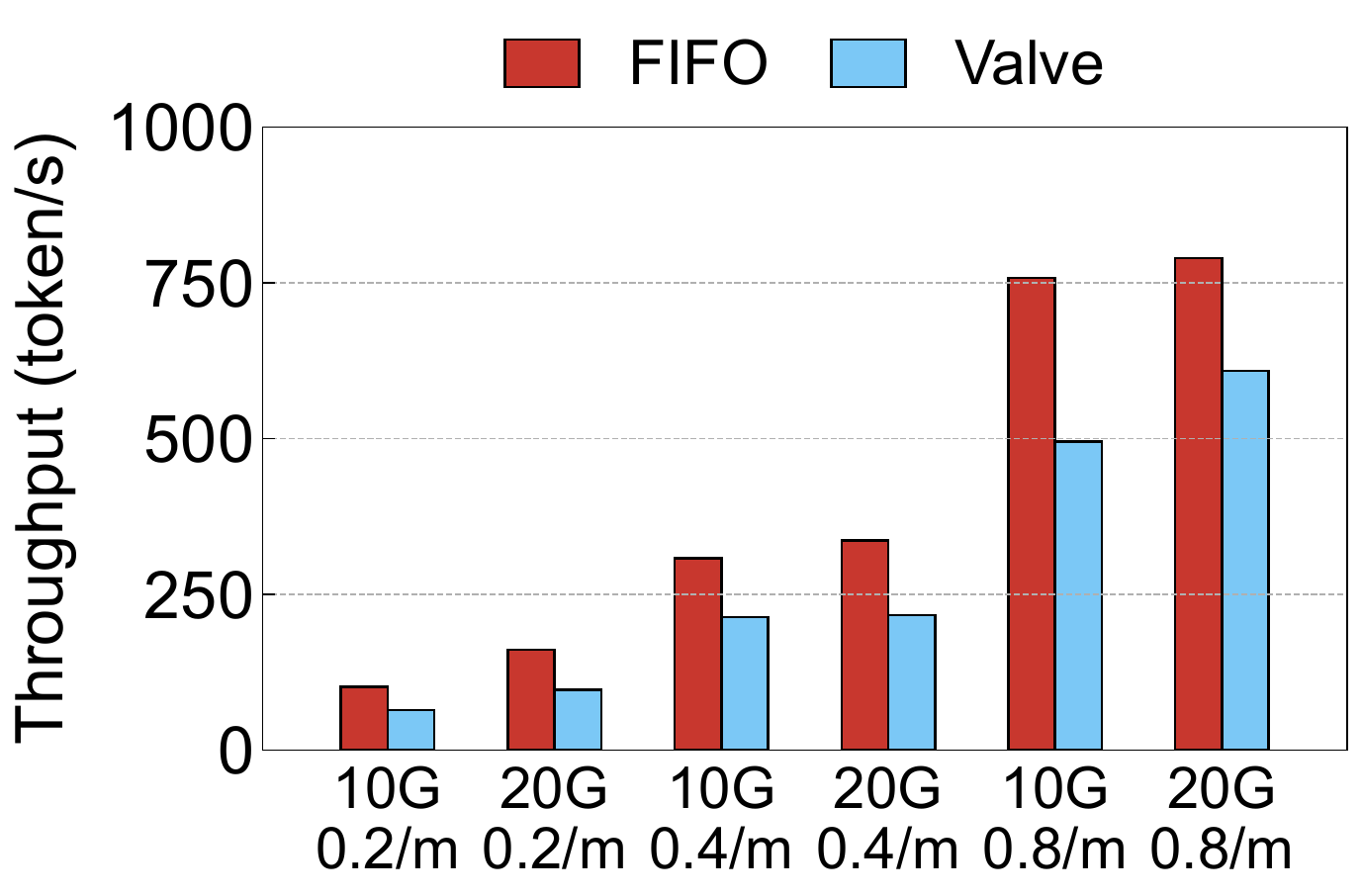}
    \caption{Effectiveness of the eviction selection.}
    \label{fig:eval3n}
\end{figure}

\parabf{Effectiveness of \sysname eviction policy.}
We further evaluate our eviction policy under varying reclamation rate and
reclaimed memory size. We compare against a FIFO baseline,
which evicts offline KV cache blocks in first-in-first-out order.
We colocate a 7B online model with a 7B offline model. As shown in
Figure~\ref{fig:eval3n}(a), by targeting blocks tied to fewer in-flight offline requests,
our policy consistently reduces throughput loss by 22.9\%--40.1\% over FIFO.

% \subsection{Effectiveness of Cluster-level Scheduler}
% \label{sec:eval:scheduler}

% \paraf{Baseline strategies.}
% We compare \sysname's cluster-level scheduler against two alternatives:
% MCF (Mem+Compute Feasible) models the memory allocated $M_{\text{req}}$ of each offline workload,
% and picks a node that continuously satisfies $M_{\text{req}}$
% and has idle GPU time fraction above the throughput threshold;
% WF (Window Feasible) models throughput with minimum memory across
% a time window and average idle GPU time fraction, and picks among nodes whose estimated throughput meets the throughput SLA.

% \parabf{Results.}
% Figure~\ref{fig:eval3n}(b) summarizes improved GPU utilization and saved GPU cards
% under the scheduling strategies.
% \sysname consistently achieves X\% higher utilization and Y greater GPU saved than the baselines.
% The gains come from \sysname's comprehensive performance model and smart scheduling,
% leading to effective workload--node placements.

% TODO: maybe add discussion on multi-resource.

% \parabf{Throughput SLA prediction accuracy.}
% Figure~\ref{fig:eval3n}(a) reports the error of our offline SLA throughput
% prediction. The error is within 5\%, with a mean of 2.1\%, ensuring precise offline SLAs.

% \input{sections/discussion}
\section{Related Work}
\label{sec:related}

\parabf{Autoscaling.}
Serverless autoscaling systems (e.g., BlitzScale, ServerlessLLM, and HydraServe) unload model weights when idle and reload them on demand~\cite{zhang2025blitzscale, lou2025towards, fu2024serverlessllm}.
This reduces steady-state GPU memory footprint, but the on-demand reload can introduce cold-start delays (e.g., TTFT spikes) under highly bursty traffic, making it hard to satisfy strict latency SLAs.
These approaches are complementary to \sysname and are suitable when bursts are milder or SLOs are less stringent.

\parabf{Multiplexing.}
Several systems multiplex multiple models on one GPU via temporal or spatial sharing~\cite{mps,li2023alpaserve,duan2024muxserve,xiang2025aegaeon,jhaqlm,nvidia_mig,ghodrati2020planaria}.
They typically target relaxed or best-effort SLAs: under contention, workloads queue for compute and memory, leading to higher latency.
% In addition, fitting multiple models often tightens per-model memory budgets (e.g., KV cache), which can further hurt tail latency.
\sysname instead targets online--offline colocation by jointly bounding interference latency and rate.

\parabf{LLM inference systems.}
Prior work improves online LLM serving via scheduling and memory management~\cite{vllm,agrawal2023sarathi,sun2024llumnix,sheng2024fairness}.
These systems focus on optimizing the serving stack (e.g., batching, KV-cache efficiency, and tail-latency control) and are orthogonal to them by providing bounded-interference for online--offline inference colocation.

\parabf{Cluster scheduling.}
Interference-aware CPU schedulers~\cite{borg,schwarzkopf2013omega,delimitrou2013paragon,delimitrou2014quasar} and ML schedulers~\cite{xiao2018gandiva,gu2019tiresias,xiao2020antman,crankshaw2017clipper,gujarati2020serving,crankshaw2020inferline} improve cluster utilization.
\sysname is orthogonal by solving new challenges in inference colocation for high utilization.

\section{Conclusion}
\label{sec:conclusion}

We present \sysname, a production-friendly online--offline colocation system that jointly bounds
\emph{preemption latency} and \emph{preemption rate}. \sysname combines channel-controlled compute
isolation with safe, rate-limited sub-layer memory reclamation, achieving
both low preemption latency and rate. Deployed on 8,054 GPUs, \sysname improves average cluster
utilization by 34.6\% and saves 2,170 GPUs, while incurring $<5\%$ TTFT increase and $<2\%$ TPOT
increase across workloads.
\label{lastpage}

\bibliography{xin}
\bibliographystyle{icml2026}

% \clearpage
% \input{sections/appendix}

\end{document}